\documentclass{report}
\usepackage{amsmath}
\usepackage{amssymb}
\usepackage{graphicx}

\begin{document}

\setlength{\textheight}{20cm}
\setlength{\textwidth}{13.5cm}
\renewcommand{\abstract}[1]{{ \footnotesize \noindent {\bf Abstract} #1 \\}}
\renewcommand{\author}[1]{\subsubsection*{\it#1}}
\newcommand{\address}[1]{\subsubsection*{\it#1}}
\newcommand{\mch}{M$_{\rm ch}$}

\def\aap{Astron.\ Astrophys.}
\def\aj{Astron.\ J.\ }
\def\apj{Astrophys.\ J.\ }
\def\apjl{Astrophys.\ J.\ Lett. \ }
\def\apjs{Astrophys.\ J.\ Suppl. \ }
\def\araa{Ann.\ Rev.\ Astron.\ Astrophys.\ }

\chapter*{Models of Type Ia Supernova Explosions}
\author{J.C. Niemeyer, M. Reinecke and W. Hillebrandt}
\address{Max-Planck-Institut f\"ur Astrophysik,
Karl-Schwarzschild-Str. 1, 85748 Garching, Germany}

\abstract{Type Ia supernovae have become an indispensable tool for
studying the expansion history of the universe, yet our understanding
of the explosion mechanism is still incomplete. We describe the
variety of discussed scenarios, sketch the most relevant physics, and
report recent advances in multidimensional simulations of
Chandrasekhar mass white dwarf explosions.}
 
\section{Introduction}

The comparison of the luminosity distances of low and high redshift
samples of type Ia supernovae (SNe Ia) is the central pillar of the
claim that our universe is accelerating \cite{RFCe98,PAGe99}. The observational
techniques, data analysis (including the light curve shape
corrections), and potential sources of systematic errors  
have been discussed and reviewed (e.g. \cite{NT00,L01}) extensively in the
literature. One, perhaps the most important, remaining uncertainty is the
possible evolution of the supernova sample. Some or even all of the
observed dimming of SNe Ia at $z \approx 1$ may still be the result of an
intrinsic trend of peak brightness with cosmic age that is not, or
only incompletely, accompanied by the corresponding change of the
light curve width. Empirically, all of the observed correlations of supernova
brightness with stellar population seem to disappear after the light curve
shape correction, hence no such trend is evident \cite{FR99}. On the other hand,
it is very difficult to extrapolate this result to higher redshifts
without a solid theoretical understanding of the origin of these
correlations and of the physics of the brightness-decline relation
itself. This, in turn, requires the construction of self-consistent
explosion models with as few adjustable parameters as possible. 

From the point of view of cosmology, in particular the planned use of
SNe Ia as high-precision tools to map out the equation of state of the
universe, some of the most urgent questions that supernova
theorists and observers must answer are: a) What are the progenitors?, b)Is
there only one class of explosions or many?, c) What is the physics
that governs the diversity of SNe Ia?, and d) How robust are the
correlations between peak brightness, light curve shape, and spectral
features with respect to, say, multidimensional mixing? This is a
challenging program that, apart from cosmology, involves a lot of
fascinating physics. 

Much progress has been made in recent years in the field of
multidimensional explosion models. We give a very brief overview of
the currently discussed explosion scenarios in Sec.~(\ref{scenarios})
(for more details and references, see \cite{HN00})
and describe our approaches to simulating Chandrasekhar mass deflagration
models in two and three dimensions in Sec.~(\ref{2/3D}).

\section{Explosion Scenarios}
\label{scenarios}

The lack of hydrogen and presence of silicon in SN Ia spectra, the
rate of the light curve decline powered by decaying nickel, and the
inferred minimum age of some SN Ia progenitors are all consistent with
thermonuclear explosions of C+O white dwarf stars \cite{HF60}. In order to
trigger the explosion, the star is believed to accrete matter from a
binary companion until critical conditions are reached. The various
explosion scenarios below differ mainly in the rate and
composition of the accreted material, the mass of the white dwarf when
it explodes, the location of the ignition, and the propagation mode
of the burning front.

The strong temperature dependence of the nuclear reaction
rates, $\dot S \sim T^{12}$ at $T \approx 10^{10}$ K,
confines the nuclear burning to microscopically thin layers that propagate
either conductively as subsonic deflagrations (``flames'') or by shock
compression as supersonic detonations. Both modes are linearly unstable
to spatial perturbations. In the nonlinear regime, the burning fronts are either
stabilized by forming a cellular structure or become fully turbulent
-- either way, the total burning rate increases as a result of flame
surface growth. Neither flames nor detonations
can be resolved in explosion simulations on stellar scales and
therefore have to be represented by numerical models.

\subsection{Chandrasekhar Mass Explosion Models}

Given the overall homogeneity of SNe Ia, the
good agreement of parameterized 1D models with observed spectra
and light curves, and their reasonable nucleosynthetic yields, the
bulk of normal SNe Ia is generally assumed to consist of exploding C+O white
dwarfs that have reached the Chandrasekhar mass, \mch, by accretion of
hydrogen or helium that burns stably to carbon and
oxygen. Flame ignition takes place near the center following roughly $10^3$
years of core convection. There is no clear identification of natural
progenitor systems, but supersoft X-ray sources (SSXS) look relatively
promising \cite{L99}. 

\subsubsection{Prompt detonation}
\label{prompt}

The first hydrodynamical simulation of an exploding \mch-white dwarf
\cite{A69} assumed that the thermonuclear combustion commences as a
detonation wave, consuming the entire star at the speed of
sound. Given no time to expand prior to being burned, the C+O material
in this scenario is transformed almost completely into iron-peaked
nuclei and thus fails to produce significant amounts of intermediate
mass elements, in contradiction to observations. It
is for this reason that \mch-explosions are believed to begin
in the deflagration (flame) mode.

\subsubsection{Pure turbulent deflagration}
\label{turbulent}

Once ignited, a subsonic thermonuclear flame
becomes highly convoluted as a result of turbulence produced by the
Rayleigh-Taylor instability (buoyancy) and the secondary Kelvin-Helmholtz
instability (shear) along the flame front. It continues to 
burn through the star until it either transitions into a detonation or
is quenched by expansion. 

By far the most simulations to date are spherically symmetric,
ignoring the multidimensionality of the flame and treating the
turbulent flame speed  $S_{\rm t}$ as a free parameter. These
studies essentially agree that 
good agreement with the observations is obtained if $S_{\rm t}$
accelerates up to roughly 30 \% of the sound speed.

In multiple dimensions, the problem of simulating turbulent
deflagrations has two aspects: the
representation of the thin, propagating surface separating hot and
cold material with different densities, and the prescription of the
local propagation velocity $S_{\rm t}(\Delta)$ of this surface as a
function of the hydrodynamical state of the large-scale calculation
with numerical resolution $\Delta$. Our solution to this problem is
sketched in Sec.~(\ref{2/3D}); for a different approach see \cite{K01}.

Most authors agree that the turbulent flame speed decouples from
microphysics on large enough scales and becomes dominated by
essentially universal hydrodynamical effects, making the scenario
intrinsically robust. A noteworthy exception is the location and
number of ignition points that significantly influences the explosion
outcome and may be a possible candidate for the mechanism giving rise
to the explosion strength variability. Other possible sources of
variations include the ignition density and the accretion rate of the
progenitor system. All of these effects may
potentially vary with composition and metallicity and can therefore
account for the dependence on the progenitor stellar population.

\subsubsection{Delayed detonation}
\label{delayed}

Turbulent deflagrations can sometimes be observed to undergo
spontaneous transitions to detonations (deflagration-detonation
transitions, DDTs) in terrestrial combustion experiments. It was suggested 
that DDTs may occur in the late phase of a \mch-explosion, providing
an elegant explanation for the initial slow burning required to
pre-expand the star, followed by a fast combustion mode that produces
large amounts of high-velocity intermediate mass elements
\cite{K91a,WW94a}. Many 1D simulations have meanwhile demonstrated
the capability of the delayed detonation scenario to provide good
fits to SN Ia spectra and light curves, as well as
reasonable nucleosynthesis products. In the best fit models, the
initial flame phase 
has a  rather slow velocity of roughly one percent of the sound speed
and transitions to detonation at a density of $\rho_{\rm DDT} \approx
10^7$ g cm$^{-3}$. The transition density was also
found to be a convenient parameter to explain the observed sequence of
explosion strengths \cite{HK96}.

On the minus side, obtaining a DDT in unconfined media without walls
or obstacles relies on local quenching of the thermonuclear burning
in a region that is many orders of magnitude larger than the flame
thickness \cite{NW97,KOW97}. This is difficult to achieve by turbulent strain
alone \cite{N99}.  Furthermore, multidimensional simulations indicate
that the turbulent flame speed is closer to 30 \% than 1 \% of the
speed of sound, so that delayed detonations seem no longer to be
needed from the energetic point of view (although they may be
beneficial for removing unburned material near the core) \cite{K01,RHN02}. 

In another variety of the delayed detonation scenario, the first
turbulent deflagration phase fails to release enough energy to unbind
the star which subsequently pulses and triggers a detonation upon
recollapse (``pulsational delayed detonation''). All pulsational
models are in conflict with current multidimensional 
simulations that predict an unbound star after the first deflagration
phase.

\subsection{Sub-Chandrasekhar Mass Models}
\label{subchandra}

C+O white dwarfs below the Chandrasekhar mass do not reach the
critical density and temperature for explosive carbon burning in the
core and need to be ignited by an external
trigger. Detonations in the accreted He layer were suggested to drive
a strong enough shock into the C+O core to initiate a secondary carbon
detonation. These so-called edge-lit detonations might explain the
class of very weak, subluminous explosions such as SN 1991bg. They are
favored mostly by the statistics of possible SN Ia 
progenitor systems \cite{YL98} and by the straightforward
explanation of the one-parameter strength sequence in terms of the
white dwarf mass. However, their ejecta structure is
characterized almost inevitably by an outer layer of high-velocity Ni
and He above the intermediate mass elements and the inner Fe/Ni core.
Therefore, these models appear to disagree photometrically and
spectroscopically with observations, but more work on the explosion
physics is needed to come to a final conclusion.

\subsection{Merging White Dwarfs}
\label{merging}

The merging white dwarf (or ``double degenerate'') scenario has to
overcome the 
crucial problem of avoiding accretion-induced collapse before it can
be seriously considered as a SN Ia candidate. If the accretion rate of
C+O onto the remaining white dwarf is larger than a few times $10^{-6}$ 
M$_\odot$ yr$^{-1}$, the most likely outcome is off-center carbon
ignition leading to an inward propagating flame that converts the star
into O+Ne+Mg. This configuration,
in  turn, is gravitationally unstable owing to electron capture onto
$^{24}$Mg and will undergo accretion-induced collapse to form a
neutron star. Dimensional analysis of the expected turbulent viscosity
suggests that it is very difficult to avoid such high accretion rates
\cite{ML90}. 

Its key strengths are a
plausible explanation for the progenitor history yielding reasonable
predictions for SN Ia rates, the straightforward explanation of the
absence of H and He in SN Ia spectra, and the existence of a simple
parameter for the explosion strength family (i.e., the mass of the
merged system).

\section{Multidimensional \mch-Models}
\label{2/3D}

In this section, we describe multidimensional simulations of what
we consider the best model for the majority of SN Ia events, i.e. the pure
turbulent deflagration model of Sec.~(\ref{turbulent}) (see
\cite{RHN02} for details). Our basic
assumptions are as follows: the initial model is a cold white dwarf
with the Chandrasekhar mass, consisting of equal amounts of carbon and
oxygen. Flame ignition starts in the
inner $\sim 150$ km of the star and the initial flame geometry acts as our
principal free
parameter. No deflagration-detonation-transition is assumed to occur.

\subsection{Modeling of the turbulent combustion front}

    The initial mixture consists of $^{12}$C and $^{16}$O at low
      temperatures. Because of the electron degeneracy the fuel
      temperature is nearly decoupled from the rest of the
      thermodynamic quantities, and since temperature is not used to
      determine the initial reaction rates, its exact value is
      unimportant.
    
When the flame passes through the fuel, carbon and oxygen are
      converted to ash, which has different compositions depending on
      the density of the unburned material. At high densities
      a mixture of $^{56}$Ni
      and $\alpha$-particles in nuclear statistic equilibrium (NSE) is
      synthesized. At lower densities burning only produces
      intermediate mass elements, which are represented by
      $^{24}$Mg. Once the density drops below $10^7$\,g\,cm$^{-3}$, no
      burning takes place.

     In the material burned to NSE, the proportion of $^{56}$Ni 
      and $\alpha$-particles changes depending on density and
      temperature even after the flame has processed the material.

     The transition densities from NSE to incomplete burning, as well as 
from incomplete burning to flame extinction were derived from data of a W7 run
provided by K.\ Nomoto. This approach is rather phenomenological, and
since these densities can have a potentially large impact on the simulation
outcome it will have to be re-examined in a thorough manner. 

  The numerical representation of the thermonuclear reaction front (i.e.\ the
  location where the ``fast'' reactions take place) is described in
  detail in \cite{RHNe99}.
  The flame front is associated with the zero level
  set of a function $G(\vec r, t)$, whose temporal evolution is given by
  \begin{equation}
     \frac{\partial G}{\partial t}
      = - (\vec{v}_u+s_u\vec{n})(-\vec{n}|\vec{\nabla}G|)\text{,}
      \label{gprop}
  \end{equation}
  where $\vec{v}_u$ and $s_u$ denote the fluid and flame propagation velocity
  in the unburned material ahead of the front, and $\vec n$ is the front normal
  pointing towards the fuel. In our case, $s_u$ identified with by the
  effective turbulent flame speed on the scale of the grid resolution,
  $S_{\rm t}(\Delta)$ (see below).
  The advection of $G$ caused by the fluid motions is
  treated by the piecewise parabolic method
  which is also used by our code to integrate
  the Euler equations. After each time step, the front is additionally 
  advanced by $s_u \Delta t$ normal to itself.

  This equation is only applied in the close
  vicinity of the front, whereas in the other regions $G$ is adjusted such
  that
  \begin{equation}
    |\vec \nabla G| = 1\text{.}
  \end{equation}

  The source terms for energy and composition due to the fast thermonuclear
  reactions in every grid cell are determined as follows:
  \begin{align}
    X'_{\text{Ashes}} &= \text{max}(1-\alpha, X_{\text{Ashes}}) \label{newash} \\
    X'_{\text{Fuel}} &= 1-X'_{\text{Ashes}} \\
    e'_{\text{tot}} &= e_{\text{tot}} + q (X'_{\text{Ashes}}-X_{\text{Ashes}})\text{,}
   \end{align}
  where $\alpha$ is the volume fraction of the cell occupied by unburned
  material; this quantity can be determined from the values of $G$ in the
  cell and its neighbors. The quantity $q$ represents the specific energy
  release of the total reaction.

  All multidimensional simulations of exploding white dwarfs share the
  problem that it is impossible to resolve all hydrodynamically unstable
  scales. The consequence is that the simulated thermonuclear flame can
  only develop structures on the resolved macroscopic scales, while the real
  reaction front will be folded and wrinkled on much finer
  scales. Simply neglecting the surface increase
  on sub-grid scales would lead to an underestimation of the energy generation
  rate, which is not acceptable; therefore a model for a turbulent
  flame speed  $S_{\rm t}$ is required to compensate this effect.

  For the case of very strong turbulence it has been shown that  $S_{\rm t}$
  decouples from the laminar flame speed and is proportional to the
  turbulent velocity fluctuations $v'$. In our simulations, $v'(\Delta)$
  is determined by using the sub-grid model introduced to SN Ia simulations by
  \cite{NH95a}. For the presented calculations a few
  corrections were applied (see \cite{RHN02}).

\subsection{Two-dimensional resolution study}
\label{calc2d}

To study the robustness of our code with respect to a change of the numerical
resolution, simulations
were performed with grid sizes of $128^2$, $256^2$, $512^2$ and $1024^2$
cells, whose corresponding resolutions in the uniform inner part of the grid
were $2\cdot10^6$\,cm, $10^6$\,cm, $5\cdot10^5$\,cm and $2.5\cdot10^5$\,cm.
The initial flame geometry (called c3\_2d) used for all these calculations
is identical to
the setup C3 presented by \cite{RHN99}: the matter
within a radius of $1.5\cdot10^7$cm from the stellar center was incinerated,
and the surface of the burned region was perturbed to accelerate the
development of Rayleigh-Taylor instabilities. 

  \begin{figure}[tbp]
    \centerline{\includegraphics[width=0.9\textwidth]{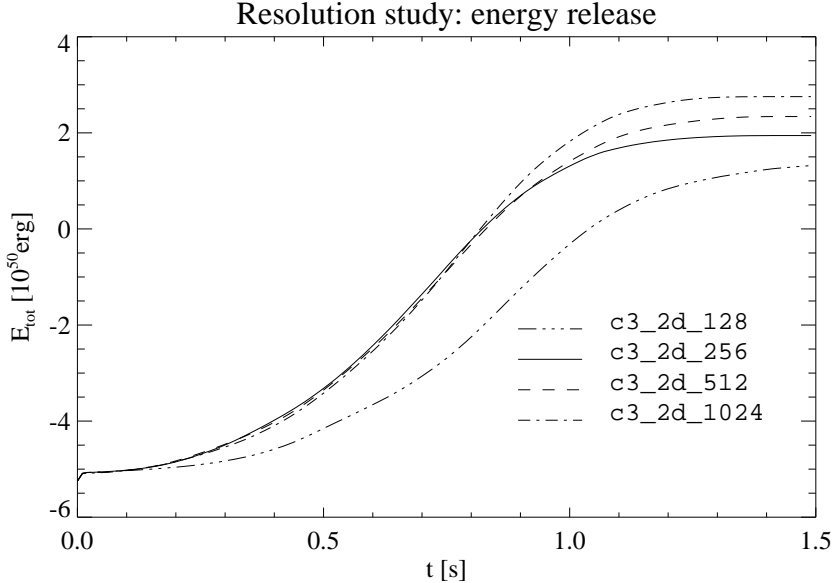}}
    \caption{Time evolution of the total energy for the initial flame geometry
             c3\_2d and different resolutions.
    During the early and intermediate explosion stages there is excellent
    agreement between the better resolved simulations.}
    \label{e_compare}
  \end{figure}
Fig.\ \ref{e_compare} shows the energy release of the models; except for
the run with the lowest resolution, the curves are nearly identical in the
early and intermediate explosion stages. Simulation c3\_2d\_128 exhibits
a very slow initial energy increase and does not reach the same final level
as the other models. Most likely this is due to insufficient resolution,
which leads to a very coarsely discretized initial front geometry and thereby
to an underestimation of the flame surface. From this result it can be deduced
that all supernova simulations performed with our code should have a central
resolution of $10^6$\,cm or better.

Overall, our model for the turbulent flame speed appears to compensate the lack
of small structures in the front very well.

\subsection{Three-dimensional simulation}
\label{calc3d}
In order to compare two- and three-dimensional simulations directly, a 3D
calculation was performed using the same initial conditions as given in
Sect.\ \ref{calc2d}.
For this purpose the initial two-dimensional flame location
was rotated by 90 degrees around the
$z$-axis and mapped onto the three-dimensional Cartesian grid consisting
of $256^3$ cells with a central resolution of $10^6$\,cm. Only one octant
of the white dwarf was simulated and mirror symmetry was assumed with respect
to the coordinate planes.

  \begin{figure}[tbp]
    \centerline{\includegraphics[width=0.8\textwidth]{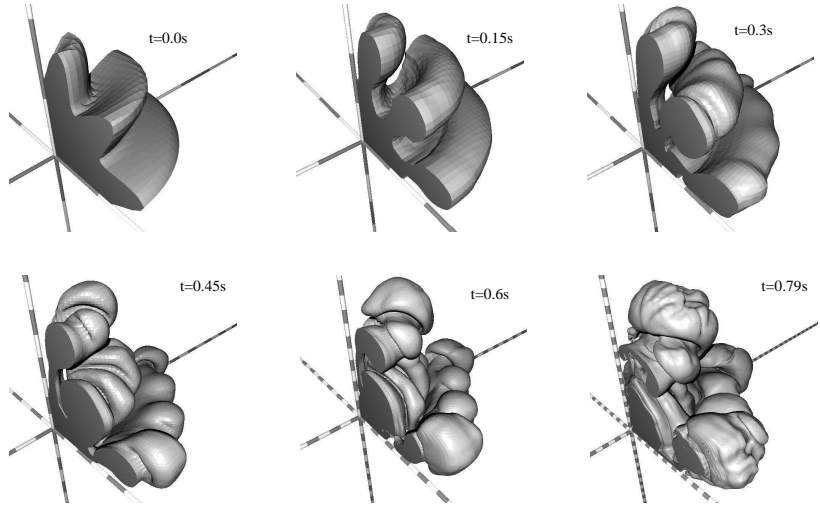}}
    \caption{Snapshots of the flame front for a centrally ignited
       three-dimensional scenario. One ring on the coordinate axes corresponds
       to $10^7$cm.}
    \label{c3_3d_front}
  \end{figure}
\begin{figure}[tbp]
    \centerline{\includegraphics[width=0.8\textwidth]{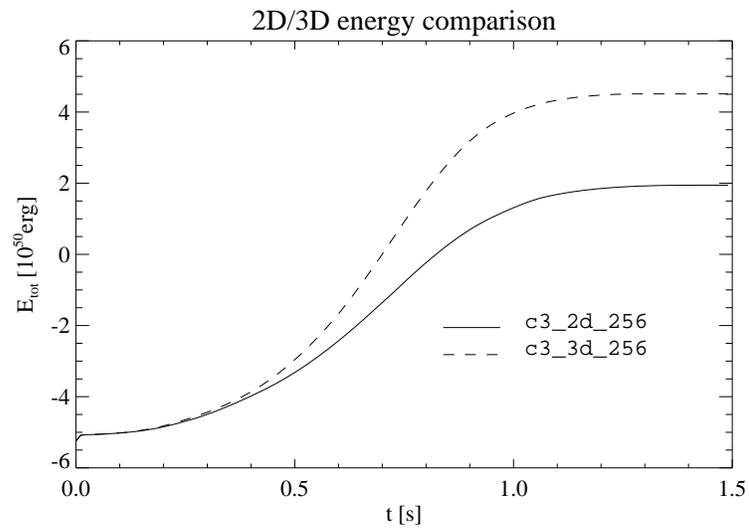}}
    \caption{Comparison of the explosion energy for identical initial conditions
     and resolution in two and three dimensions. After the loss of axial
     symmetry (at $t\approx0.3$\,s) the larger flame surface in the
     three-dimensional model leads to more vigorous burning.}
    \label{comp2d3d}
  \end{figure}

The initial configuration, as well as snapshots at later times, are shown
in Fig.\ \ref{c3_3d_front}. Obviously, the initial axisymmetry is lost
after 0.2\,--\,0.3\,s, although
  no explicit perturbation in $\varphi$-direction was applied to the front.
  This happens because the initial flame geometry cannot be mapped perfectly
  onto a Cartesian grid and therefore the front is not transported with exactly
  the same speed for all $\varphi$. During the next few tenths of a
    second, these small deviations cause the 
  formation of fully three-dimensional Rayleigh-Taylor-mushrooms, leading to a
  strong convolution of the flame. As expected, this phenomenon
  has a noticeable influence on the explosion energetics; this is illustrated
  in Fig.\ \ref{comp2d3d}.
  Before the loss of axial symmetry in c3\_3d\_256, the total energy evolution
  is almost identical for both simulations, which strongly suggests that the
  two- and three-dimensional forms of the employed turbulence and level set
  models are consistent, i.e.\ that no errors were introduced during the
  extension of these models to three dimensions. In the later phases the 3D
  model releases more energy as a direct consequence of the surface increase
  shown in Fig.\ \ref{c3_3d_front}.

\subsection{Discussion and conclusions}
\label{discuss}

The white dwarf becomes unbound in all of our models, which implies that no
recontraction (and hence no pulsational detontation) occurs. Nevertheless only 
the three-dimensional model results in a powerful enough explosion to qualify as a
typical SN Ia; the two-dimensional scenarios are too weak to accelerate
the ejecta to the speeds observed in real events and produce too little nickel
to power a standard SN Ia light curve.

The 3D calculation is a good candidate
for typical SN Ia explosions, at least with respect to explosion strength and
remnant composition. The produced nickel mass of 0.53 M$_\odot$ falls
well into the range of $\approx$\,0.45\,--\,0.7\,M$_{\odot}$ determined by
\cite{CLV00} for several typical events, 
and it can be deduced from the amount of 0.18  M$_\odot$ of
``magnesium'' in the ejecta that enough intermediate mass elements
were synthesized to explain the observed spectral features.

Qualitatively, our results for the explosion energetics are in relatively good
agreement with recent simulations performed by \cite{K01}, employing 
different numerical models and initial conditions. We interpret this
as an inherent robustness of the Chandrasekhar mass deflagration
scenario. 

This development is a major step towards constructing self-consistent
models for type Ia supernovae. The remaining free parameters are
chosen according to our best understanding of the unresolved
physics without any reference to obtaining ``good'' explosions. As
there will always be relevant unresolved scales in this 
problem, we need to keep improving our understanding by performing
numerical experiments of turbulent thermonuclear combustion on
microscopic and intermediate scales. Nevertheless, our results make us
optimistic that multidimensional models will soon allow us to
understand how type Ia supernovae really work.

\bigskip
 
\noindent {\Large{\bf Acknowledgements}}

\bigskip

We thank Bruno Leibundgut and Paolo Mazzali for helpful discussions.

\end{document}